\documentclass[a4paper]{article}
\usepackage{color}
\usepackage{amsthm}
\usepackage{amsmath}
\usepackage{graphicx}
\usepackage{amssymb}
\usepackage{bbold}
\usepackage{esint}
\usepackage{subfigure}
\usepackage{color}
\usepackage[all]{xy}

\usepackage{anysize}
\marginsize{2cm}{2cm}{2cm}{2cm}

\def\beq{\begin{equation}}
\def\eeq{\end{equation}}
\def\bea{\begin{eqnarray}}
\def\eea{\end{eqnarray}}

\def\l{\left(}
\def\r{\right)}

\makeatletter

\theoremstyle{plain}

  \theoremstyle{remark}

\begin{document}

\title{Generating domain walls between topologically ordered phases using quantum double models}
\author{Miguel Jorge Bernab\'{e} Ferreira,$^{a}$\footnote{migueljb@if.usp.br} ~Pramod Padmanabhan,$^{a,b}$\footnote{pramod23phys@gmail.com}~\\ Paulo Teotonio-Sobrinho$^{a}$\footnote{teotonio@fma.if.usp.br}  } 
\maketitle

\begin{center}
{\small $^{a}$ Departamento de F\'{i}sica Matem\'{a}tica, Universidade de S\~{a}o Paulo}

{\small  Rua do Mat\~ao Travessa R Nr.187, S\~ao Paulo, CEP 05508-090}

{\small $^{b}$ Department of Physics and Astronomy, University of California, Irvine}

{\small 4129 Frederick Reines Hall Irvine, CA 92697-4575}
\end{center}

\begin{abstract}

Transitions between different topologically ordered phases have been studied by artificially creating boundaries between these gapped phases and thus studying their effects relating to condensation and tunneling of particles from one phase to the other. In this work we introduce exactly solvable models which are similar to the quantum double models (QDM) of Kitaev where such domain walls are dynamically generated making them a part of the spectrum. These systems have a local symmetry and may or may not have a global symmetry leading to the possibility of a ground state degeneracy arising from both, global symmetry breaking and a topological degeneracy. The domain wall states now separate the different topologically ordered phases belonging to the different sectors controlled by the global symmetry, when present. They have interesting properties including fusion with the deconfined anyons of the topologically ordered phase, to either create new domain walls, or their annihilation. They can also act like a scatterer permuting anyons of the topologically ordered phases on either side of the domain wall. Thus these domain wall states can be thought of as a synthetic scatterer which can be created in any part of the lattice. We show these effects for the simplest case of the $\mathbb{Z}_2$ toric code phase decorated with a global $\mathbb{Z}_2$ symmetry and then make remarks about the case when we have the QDM based on two arbitrary groups $G_1$ and $G_2$ in which case we may no longer have a global symmetry.  

\end{abstract}

\section{Introduction}
~

Ideas for classifying quantum phases of matter that go beyond the description due to Landaus's symmetry breaking scheme were conceived since the discovery of the integer and fractional quantum Hall effect and high temperature superconductivity~\cite{h1, h2, h3, h4}. Since then the study of low temperature phases exhibiting topological order have been given new life due to their potential applications in fault tolerant computation~\cite{freed, nayak, kitToric} and in their description of topological insulators and their generalizations in the form of symmetry protected topological phases~\cite{bernBook, ti1, ti2, wen2}. These phases are usually described by a topological quantum field theory in the continuum. There are also examples of exactly solvable Hamiltonians describing the topological phases. In most cases these Hamiltonians are sums of commuting projectors, with the most prominent set of models being given by the quantum double models (QDM) of Kitaev~\cite{kitToric, aguado1}.

The QDM models on lattices with boundaries have been studied in the past~\cite{Shor, kk, kb, kong}. The types of boundaries considered were either smooth or rough. There are good reasons to consider models with boundary. In 3D, for example, it is not realistic to consider boundary less systems. More importantly, the relationship between boundary and bulk degrees of freedom provide us new interesting phenomena. These models are described with an exactly solvable Hamiltonian which is split into bulk and boundary parts. They exhibit different topological phases in the bulk and the boundary. In general the phase on the boundary is characterized by anyon types belonging to a subcategory of the fusion category describing the bulk phase. In~\cite{Shor} domain walls are considered separating two different topologically ordered phases governed by the IRR's of the quantum double of two different groups $G$ and $G'$. The conditions for anyons to tunnel between these two phases are studied in detail. In the special case when one of the groups become trivial the domain wall turns into a boundary, which have been analyzed in more detail in~\cite{kk} for the string-net models~\cite{LevinWen} as well. The set of anyons on the boundary are then determined by fusing the bulk anyons into the domain wall or in other words by {\it condensing} the bulk anyons on the boundary. 
In all these examples the boundary is a physical trait of the lattice where the system lives on. We may say that it is put in by hand and its location on the lattice is fixed. 

In this work we introduce exactly solvable models of the quantum double type that contain these domain walls as part of their spectrum. This means that there are operators that act on the Hilbert space of the theory that can create these domain walls locally as an eigenstate of the Hamiltonian. The reason this happens in these models is because the degrees of freedom on the local Hilbert space are no longer labeled by elements of a group as it is in the case of the QDM of Kitaev based on group algebras. Here the basis of the local Hilbert space is generated by the same elements as in the group algebra plus an extra label. In other words the model now has two degrees of freedom, namely the elements of the group algebra and an extra quantum number called $s$ belonging the to the set $s\in\{1,2\}$.  This introduces new excitations in the theory that cannot be moved around like the deconfined anyonic excitations of the QDM of Kitaev. Thus we can now create ``walls'' or ``regions'' of these excitations and we see that they are similar to the domain wall excitations studied in the past.
 
 A more physical way of looking at this is that the model considered here has both a global and local $\mathbb{Z}_2$ symmetry. Thus their ground states obtain a degeneracy from two mechanisms namely the global symmetry breaking mechanism, as it happens in say the Ising model and a topological degeneracy, when the system is placed on a surface with a non-zero genus as it happens in the toric code. Thus the global $\mathbb{Z}_2$ symmetry is broken at the ground sate level with the system settling down in either of the two toric code topological phases labeled by $s=1,2$.. The domain wall states are obtained by operating in the sector controlled by the global symmetry, namely the $s$ degrees of freedom, and thus they separate regions that are locally in one of the two toric code phases. Thus we can interpret this system as a toric code decorated with an additional $\mathbb{Z}_2$ global symmetry whereas the toric code in it's original formulation has only a local $\mathbb{Z}_2$ symmetry. Recently such gauge systems decorated with a global symmetry have been studied under the name symmetry enriched topological (SET) phase \cite{herm1, herm2, fid}. The ground states in these systems preserve the global symmetry unlike the examples studied here. Other examples with both a global and local symmetry include the quantum Hall ferromagnets~\cite{qhf}. 

These domain wall states can manifest themselves in different forms each with interesting properties. The deconfined anyons on each of the phases can be fused into the domain walls to become a part of it. This has been termed as fusing anyons into domain wall excitations in the literature. Once we fuse anyons into the domain wall we create states which are essentially isolated excitations connected to the domain wall through a string which then act like sinks for the other anyons that condense into them just like how they condense into the physical boundaries considered in~\cite{Shor, kk, kb}. Thus these are sinks which can be created locally in any part of the lattice. Apart from this the domain walls can act as annihilators by killing states that fuse into them. This is a new feature in this model. Finally they act as a scatterer helping us tunnel between the anyons of the two topologically ordered phases. In particular they can switch between same anyon types or different anyon types. The effect is similar to the anyon symmetry of bi-layer systems.

The paper is organized as follows. Section 2 describes the setup of our system which includes the Hilbert space and the operators that go into the quantum double Hamiltonian. The spectrum is similar to the that of the QDM. The Hamiltonian is a sum of commuting projectors and thus all its eigenstates can be easily obtained. The ground states and the anyonic excitations are presented in sections 3 and 4 respectively. The domain wall states are also shown in section 4. Their effects are studied in detail in section 5. In section 6 we briefly generalize the model to the case where the two different topologically ordered phases are given by the QDM of a pair of distinct groups. Some crucial remarks make up section 7.

\section{The Hilbert space and the Hamiltonian}
~

The model is defined on a two dimensional square lattice with the local Hilbert space $\mathcal{H}_l$ on the links $l$. The degrees of freedom have an extra quantum number $s$ beyond the usual $\mathbb{Z}_2$ group elements $+1$ and $-1$ of the toric code. This extra quantum number takes values in a discrete set with two elements, $I=\{1,2\}$. Thus a basis for the local Hilbert space is given by $\{\vert s,g\rangle :g\in \mathbb{Z}_2,\;s\in I\}=\{\vert 1,+1 \rangle,\vert 1,-1 \rangle,\vert 2,+1 \rangle,\vert 2,-1 \rangle\}$\footnote{Note that the local Hilbert space $\mathcal{H}_l$ can be thought of as $\mathcal{H}_l= \mathbb{C}^2\otimes\mathbb{C}(\mathbb{Z}_2)$, where $\mathbb{C}(\mathbb{Z}_2)$ is the group algebra of $\mathbb{Z}_2$ and $\mathbb{C}^2$ is a two dimensional vector space whose basis elements are labeled by the elements of $I=\{1,2\}$. Thus the basis elements $\vert s,g\rangle$ of $\mathcal{H}_l$ can also be represented by $\vert s,g\rangle=\vert s \rangle\otimes \vert g \rangle$.
}. The full Hilbert space is given by $\mathcal{H}=\bigotimes_l \mathcal{H}_l$. A basis of $\mathcal{H}$ is given by 
$$\{\vert \underbrace{s_1,g_1}_{\tiny\hbox{link 1}}~;~\underbrace{s_2,g_2}_{\tiny\hbox{link 2}}~;~\cdots~;~\underbrace{s_{n_l},g_{n_l}}_{\tiny\hbox{link } n_l}  \rangle:g_i\in\mathbb{Z}_2,\; s_i\in I\}\;.$$

The operators acting on the Hilbert space $\mathcal{H}$ can be written out of following local operators 
\beq\sigma^{\mu\nu}:=\sigma^\mu\otimes\sigma^\nu\;,\eeq
where $\mu,\nu\in\{0,x,y,z\}$ and $\sigma^\mu$ are the Pauli matrices \footnote{We are considering the canonical basis $\vert +1 \rangle = \left(\begin{array}{c} 1 \\ 0 \end{array}\right), \; \vert -1 \rangle = \left(\begin{array}{c} 0 \\ 1 \end{array}\right),\; \vert 1 \rangle = \left(\begin{array}{c} 1 \\ 0 \end{array}\right),\; \vert 2 \rangle = \left(\begin{array}{c} 0 \\ 1   \end{array}\right)$ which are eigenvectors of $\sigma^z$.} with $\sigma^0=\mathbb{1}$.
We define two very important projectors which we will call $P^r$ with $r\in I$. Their action is given by
\beq P^r\vert s,g\rangle=\delta(r,s)\vert s,g \rangle\;.\eeq
These projectors will play a very important role in this model as we will see later and they can be easily be written as linear combination of $\sigma^{\mu\nu}$.

\begin{figure}[h!]
\centering
\subfigure[$A_v^1$ acts multiplying the gauge degrees of freedom by $-1$.]{
\includegraphics[scale=1]{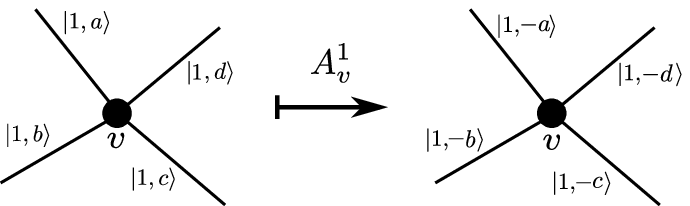}\label{Avphi-a}
}
\hspace{2cm}
\subfigure[If at least one local state around the vertex $v$ has $s=2$ the $A_v^1$ operators projects it to the zero vector.]{
\includegraphics[scale=1]{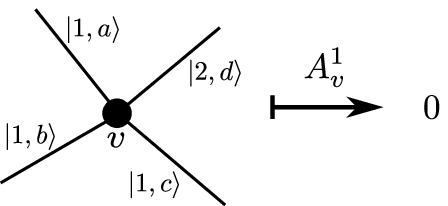}\label{Avphi-b}
}
\caption{The action of the $A_v^1$ vertex operator.}
\label{Avphi}
\end{figure}

The Hamiltonian is made of vertex and plaquette operators like the toric code Hamiltonian
\begin{equation}
\label{H}
H=-\sum_p B_p -\sum_l A_v\;.
\end{equation}
Their action are however sightly different. Consider the vertex operator $A_v$. It is convenient to write the vertex operator as a sum of two other vertex operators $A_v=A_v^1 +A_v^2$. The operator $A_v^1$ acts exactly like the usual Toric Code vertex operator flipping $g\mapsto -g$ when all the local states around the vertex $v$ has quantum number $s=1$. Otherwise, if at least one local state around the vertex $v$ has quantum number $s=2$, the vertex operator $A_v^1$ projects the state to the zero vector of $\mathcal{H}$. The action is shown in figure \ref{Avphi}.
The operator $A_v^2$ does the same for $s=2$. These operators can be realized as
\beq A_v^r = \bigotimes_{l\in\hbox{star}(v)}\Sigma_l^x P_l^r \;\;\; \hbox{with} \;\; r\in I\;,
\eeq
where $\Sigma^x=\sigma^{0x}=\mathbb{1}\otimes\sigma^x$. It is not difficult to see that the vertex operators obey the following relations for all vertices $v$ and $v^\prime$
\begin{subequations}
\begin{eqnarray}
\label{AvphiAvvarphi}A_v^r A_v^s &=& \delta(r,s) \bigotimes_{l\in\hbox{star}(v)}P_l^1\;, \\ 
\label{Avsquare}(A_v)^2 &=& \bigotimes_{l\in\hbox{star}(v)} P_l^1+\bigotimes_{l\in\hbox{star}(v)}P_l^2\;, \\ 
\left[A_v,A_{v^\prime}\right]&=& 0 \;.
\end{eqnarray}
\end{subequations}

The plaquette operator $B_p$ is defined in a similar way, it is also made up of two others plaquette operators  $B_p=B_p^1+B_p^2$, which are defined as follows
\beq \label{Bp1} B_p^r = \bigotimes_{l\in\partial p}\Sigma_l^z P_l^r \;\; \; \hbox{with}\;\; r\in I \;,\eeq
where $\Sigma^z=\sigma^{0z}=\mathbb{1}\otimes\sigma^z$. Again, in the case where all the local states living around the plaquette have quantum number $s=1$ (or $s=2$) the plaquette  operator $B_p^1$ (or $B_p^2$) gives the same state back with the eigenvalue equal to the holonomy around the plaquette $p$. And if there is a mix of two different values of $s$, the plaquette operator projects the state to the zero vector. The plaquette operator obey the following relations for all plaquettes $p$ and $p^\prime$
\begin{subequations}
\begin{eqnarray}
B_p^r B_p^s &=& \delta(r,s) \bigotimes_{l\in\partial p} P_l^1\;, \\ 
\label{Bpsquare}(B_p)^2 &=& \bigotimes_{l\in\partial p} P_l^1+\bigotimes_{l\in\partial p} P_l^2\;, \\ 
\left[B_p,B_{p^\prime}\right]&=& 0 \;.
\end{eqnarray}
\end{subequations}
Besides, the vertex and plaquette operators satisfy the following relations for all vertices and plaquettes
\begin{subequations}
\begin{eqnarray}
[A_v^r,B_p^s] &=& 0\;,\; \forall r,s\; \\ 
\label{AB-BA}\left[A_v,B_p\right] &=& 0\;.
\end{eqnarray}
\end{subequations}
In other words the Hamiltonian is a made up of a sum of commuting operators, which make it exactly solvable.

\section{The ground state}
~

\begin{figure}[h!]
\centering
\subfigure[Ground states with $s=1$.]{
\includegraphics[scale=1]{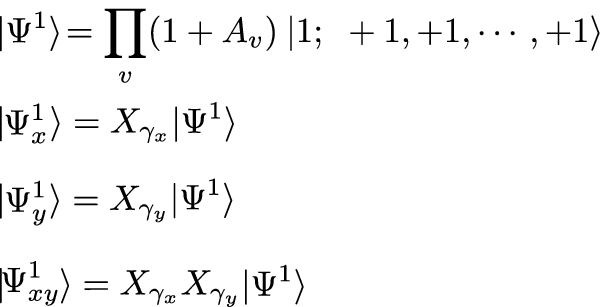}\label{TCgroundstates-a}
}
\hspace{.1cm}
\subfigure[Non-contractible loops on the torus.]{
\includegraphics[scale=1]{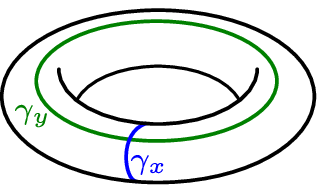}\label{TCgroundstates-b}
}
\hspace{.1cm}
\subfigure[Ground states with $s=2$.]{
\includegraphics[scale=1]{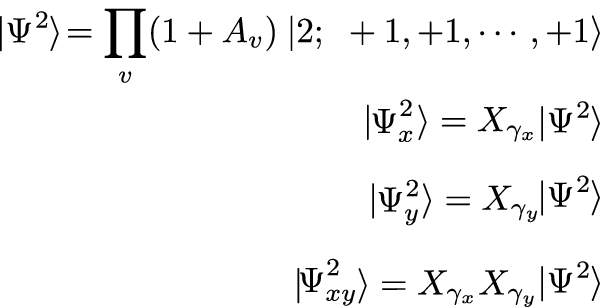}\label{TCgroundstates-c}
}
\caption{Ground states of the Hamiltonian \eqref{H}.}
\label{Bpphi}
\end{figure}

Since the Hamiltonian is a sum of commuting operators it can be easily diagonalized and just as in the toric code case, this model has degenerate ground states. The vertex and plaquette operators obey the properties \eqref{Avsquare} and \eqref{Bpsquare} which means their eigenvalues are either $-1$, $0$ or $+1$. Due to this fact and equation \eqref{AB-BA} the vacuum subspace $\mathcal{H}^0$ is defined as
$$\mathcal{H}^0=\{\vert \Psi^0 \rangle\in \mathcal{H}:B_p\vert \Psi^0 \rangle=A_v\vert \Psi^0 \rangle=\vert \Psi^0 \rangle\;, \forall p,v \}\;.$$
Consider only states with $s=1$ on all the links, which we will represented by $\vert 1;~g_1,g_2,\cdots,g_{n_l}\rangle$. Once restricted to this subspace of $\mathcal{H}$, the Hamiltonian \eqref{H} is exactly like the toric code Hamiltonian, and one  ground state is the following
$$\vert \Psi^1 \rangle = \prod_v\left(\mathbb{1}+A_v \right)\vert 1;~+1,+1,\cdots,+1 \rangle\;.$$
Using the relations \eqref{AvphiAvvarphi} and \eqref{AB-BA} it is not difficult to see that the above state satisfies $B_p\vert \Psi^1 \rangle=A_v\vert \Psi^1 \rangle=\vert \Psi^1  \rangle\;,\;\forall p,v$. There are three more ground states one can obtain from $\vert \Psi^1 \rangle$ by flipping $g\mapsto -g$ along a non contractible loop of the torus, in other words, by applying the operator 
$$X_\gamma=\bigotimes_{l\in \gamma}\Sigma_l^x\;,$$
where $\gamma$ is a non contractible loop on the dual lattice of a torus. Since there are only two non contractible loops on the torus (up to homotopic deformations) the four ground states one can obtain are the ones represented in figure \ref{TCgroundstates-a}.
Likewise, if we do the same analysis for the subspace of $\mathcal{H}$ made up of only linear combinations of vectors of the kind $\vert 2;~g_1,g_2,\cdots,g_{n_l}\rangle$, with $s=2$ on all the links, we will find out four more states which are also ground states of the Hamiltonian. These states are represented in figure \ref{TCgroundstates-c}. An important thing is that the states with $s=1$ cannot be mapped into the ones with $s=2$ by gauge transformations (vertex operator action), thus making the ground state degeneracy for this model twice the toric code ground state degeneracy. This can also been seen due to the phenomena of spontaneous symmetry breaking of the global $\mathbb{Z}_2$ symmetry. A basis for the subspace $\mathcal{H}^0$ is the following: $\{\vert \Psi^1 \rangle,\vert \Psi_x^1 \rangle,\vert \Psi_y^1 \rangle,\vert \Psi_{xy}^1 \rangle,\vert \Psi^2 \rangle,\vert \Psi_x^2 \rangle,\vert \Psi_y^2 \rangle,\vert \Psi_{xy}^2 \rangle\}$.

\section{Excited states}
~

The excitations of these model are similar to the toric code but with important differences according to the values of $s$. The elementary excitations are charges and fluxes but they now come in two kinds according to the $s$ label they carry. The operators which creates excitations are string operators which act on either a path $\gamma$ on the direct lattice or a path $\gamma^*$ on the dual lattice. In the following let us study each of these operators separately.

\subsection{Charge and fluxes: toric code-like anyons}
~

\begin{figure}[h!]
\centering
\subfigure[Toric code-like charge excitations.]{
\includegraphics[scale=1]{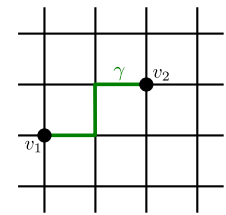}\label{Avexc}
}
\hspace{2cm}
\subfigure[Toric code-like flux excitations.]{
\includegraphics[scale=1]{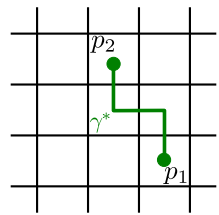}\label{Bpexc}
}
\caption{Toric code-like excitations.}
\label{exc}
\end{figure}

Consider the string operator 
$$Z_\gamma=\bigotimes_{l\in\gamma}\Sigma_l^z\;.$$
The $Z_\gamma$ operator trivially commutes with the plaquette operator and since $\Sigma^x\Sigma^z = -  \Sigma^z\Sigma^x$, the state $Z_\gamma\vert \Psi^s \rangle$, with $\vert \Psi^s \rangle$ a ground state of the Hamiltonian \eqref{H}, is an excited state which has two charge excitations on the vertices $v_1$ and $v_2$, as shown in figure \ref{Avexc}.
Let us call these charge excitations $e_s$, note they can be of two different kinds, namely $e_1$ and $e_2$, depending on the $s$ value of the ground state it was created from. The operator
$$X_{\gamma^*}=\bigotimes_{l\in\gamma^*}\Sigma_l^x\;,$$
is the string operator which creates fluxes. It trivially commutes with the vertex operators creating flux excitations on plaquettes $p_1$ and $p_2$ at the end points of $\gamma^*$, as shown in figure \ref{Bpexc}. Again, the flux excitations can be of two different kinds, namely $m_1$ and $m_2$ distinguished by the $s$ value.

All these excitations are toric code-like anyons, apart from the fact that the charges and fluxes are now of two different kinds. Since each kind of flux (charge) are created from different ground states, they can not be fused, in other words an anyon $x_s$ (with $s\in I$ and $x=e,m$) can only be fused with another anyon $y_{s^\prime}$ if $s=s^\prime$, which means there are two distinct set of anyons whose generators are $\{e_1,m_1\}$ and $\{e_2,m_2\}$. The fusion rules of such anyons are the following
\begin{subequations}
\bea
e_s\times m_s &=& \epsilon_s \;,\\
e_s\times \epsilon_s &=& m_s \;,\\
m_s\times \epsilon_s &=& e_s\;,\\
e_s \times e_s &=& m_s\times m_s = \epsilon_s \times \epsilon_s = 1_s\;,
\eea
\end{subequations}
where $\epsilon_s$ is called a dyon and $1_s$ stands for the vacuum with label $s$.

\subsection{Domain walls}
~

\begin{figure}[h!]
\centering
\subfigure[Graphical representation of a local state $\vert s,g\rangle$.]{
\includegraphics[scale=1]{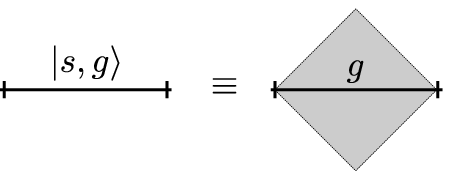}\label{grafical-a}
}
\hspace{1cm}
\subfigure[The white diamond means a local state with $s=1$.]{
\includegraphics[scale=1]{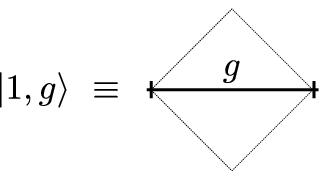}\label{grafical-b}
}
\hspace{1cm}
\subfigure[The green diamond means a local state with $s=2$.]{
\includegraphics[scale=1]{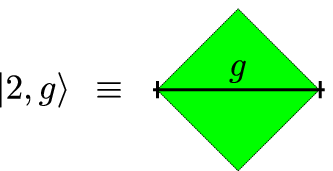}\label{grafical-c}
}
\caption{Graphical notation for different states.}
\label{grafical}
\end{figure}

As we have seen before the plaquette and vertex operators have zero as an eigenvalue apart from the usual $\pm 1$. A state which gives eigenvalue zero for a plaquette (vertex) operator is said to have a zero flux (charge) excitation localized on such a plaquette (vertex). This kind of state can be obtained by flipping the quantum number $s$ of a single link $l$ belonging to $\partial p$ (or star($v$)). An operator that does this is given by $F_l=\sigma_l^{x0}=\sigma^x\otimes\mathbb{1}$. At this point a graphical notation would be useful for understanding these excitations. For a given local state $\vert s , g\rangle$ we associate a diamond as shown in figure \ref{grafical}. These diamonds represent the $s$ quantum number. If $s=1$ we color it white as shown in figure \ref{grafical-b} or if $s=2$ we color it green as shown in figure \ref{grafical-c}.
The operator $F_l$ can be thought as the operator which flips the diamond color. The properties of such an operator are the following
\begin{subequations} 
\begin{eqnarray}
F_l P_l^r &=& P_l^{\tilde{r}} F_l\;\; \hbox{for} \; r\in I\;, \\
\left[F_l,\Sigma^x \right] &=& \left[F_l,\Sigma^z \right] = 0\;,
\end{eqnarray}
\end{subequations}
where $\tilde{r}=-r+3$.

\begin{figure}[h!]
\centering
\subfigure[Creation of a pair of zero flux and zero charge excitations.]{
\includegraphics[scale=1]{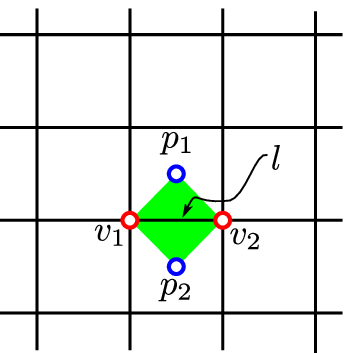}\label{domain-a}
}
\hspace{1cm}
\subfigure[Creation of zero flux and charge excitations along a path $\gamma^*$ on the dual lattice.]{
\includegraphics[scale=1]{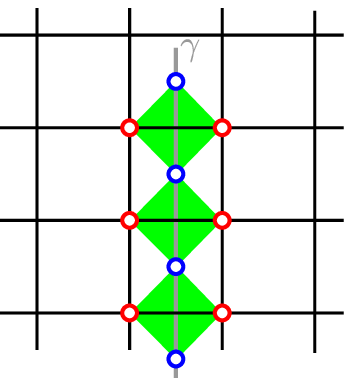}\label{domain-b}
}
\hspace{1cm}
\subfigure[Creation of a domain wall. The region with white diamond corresponds to the phase $s=1$ while the region with green diamonds corresponds to the phase $s=2$.]{
\includegraphics[scale=1]{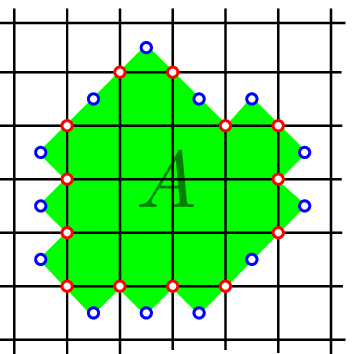}\label{domain-c}
}
\caption{Zero flux and zero charge excitations are created by flipping the quantum number $s$.}
\label{domain}
\end{figure}

Let us now see how to create the zero fluxes excitations. We act on the ground state, $\vert \Psi^1 \rangle$ written in the previous section. As we already saw $B_p\vert \Psi^1 \rangle=A_v\vert \Psi^1 \rangle=\vert \Psi^1 \rangle$ for all vertices and plaquettes. The state $F_l\vert \Psi^1 \rangle$ has now a zero flux excitation localized on each plaquette which share the link $l$, moreover it has also zero charge excitations localized on the vertices which are on the end points of link $l$. The reason for that is that the zero eigenvalues for the plaquette operators comes from the case where there are at least two diamonds of different colors sharing the same plaquette, but since the ground state $\vert \Psi^1 \rangle$ has $s=1$ for all the links the plaquette operators won't give zero as eigenvalue. However when the operator $F_l$ is applied on the link $l$, it creates a single link of different diamond color which leads to these zero fluxes excitations on all the plaquettes that share the link $l$. Likewise, the vertex operators on the end points of the link $l$ will also be excited as $F_l$ flips $A_v^1$ to an operator of the type shown in figure \ref{Avphi-b}. This operator measures the zero charge excitations. Thus the state ends up with two zero flux excitations and two zero charges excitations. Thus we have
\begin{subequations}
\begin{eqnarray}
B_{p_1}\left(F_l\vert \Psi^1 \rangle \right) &=& B_{p_2}\left(F_l\vert \Psi^1 \rangle \right)=0\;, \\
A_{v_1}\left(F_l\vert \Psi^1 \rangle \right) &=& A_{v_2}\left(F_l\vert \Psi^1 \rangle \right)=0\;, \\
B_p\left(F_l\vert \Psi^1 \rangle \right) &=& A_v\left(F_l\vert \Psi^1 \rangle \right)=F_l\vert \Psi^1 \rangle \;\;\hbox{for}\;\; p\neq p_1, p_2\;,\; v\neq v_1,v_2\;.
\end{eqnarray}
\end{subequations}
The picture in figure \ref{domain-a} illustrates this fact, showing the mismatch of diamond colors. We call the zero flux excitation $m_0$ while the zero charge excitations are called $e_0$.

If the flip operator $F_l$ is applied on the ground state $|\psi^1\rangle$ along a path\footnote{This path can be a path either on the direct or the dual lattice.} $\gamma$, that is
$$F_\gamma=\bigotimes_{l\in\gamma} F_l,$$
it will create a couple of zero flux and zero charges all along it's path and not just in the end points like in the toric code. We say that these excitations are not deconfined like the ones in the toric code. This is shown in figure \ref{domain-b}. 

We can also apply the operator $F_l$ in all the links which are inside an area $A$ (including its boundary $\partial A$). This is done through applying the following operator
$$F_A=\bigotimes_{l\in A} F_l\;.$$
In figure \ref{domain-c} one can see the effect of applying such an operator on the ground state. Note that there is no diamond color mismatch inside the area. The zero flux and zero charge excitations will appear only on its boundary as it can be seen in figure \ref{domain-c}. The energy cost to create such a state is proportional to length of the path $\partial A$. This state is a domain wall which separates the two different toric code sectors labeled by $s$. In the case of figure \ref{domain-c}, the state with $s=1$ outside the area and $s=2$ inside it. These domain walls are very important, for example they allow toric code-like excitations to tunnel between two different values of $s$ apart from other effects. These will be illustrated in what follows.

\section{Domain wall effects}
~

\begin{figure}[h!]
\centering
\subfigure[Creation of state with a domain wall out of the ground state $\vert \Psi^1 \rangle$.]{
\includegraphics[scale=1]{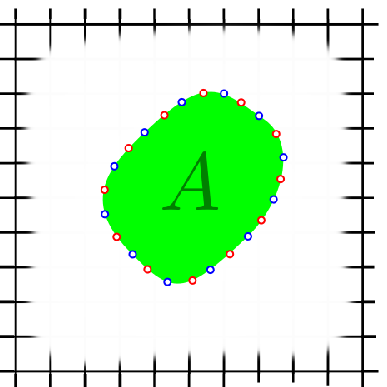}\label{domainTC-a}
}
\hspace{1cm}
\subfigure[Creation of pair of toric code-like anyons  $x_1$ with $s=1$.]{
\includegraphics[scale=1]{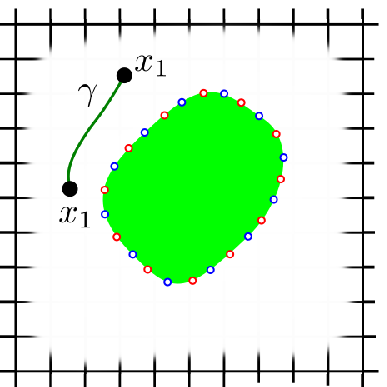}\label{domainTC-b}
}
\hspace{1cm}
\subfigure[Creation of toric code-like anyons $x_2$ with $s=2$.]{
\includegraphics[scale=1]{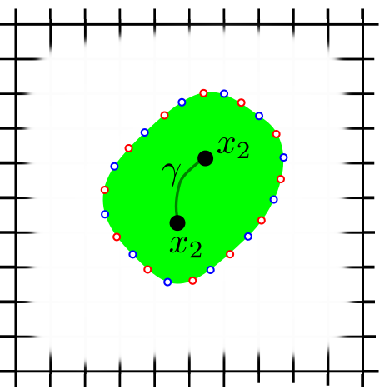}\label{domainTC-c}
}
\caption{Creation of toric code-like charge excitations of two different kinds in the background of domain walls.}
\label{domainTC}
\end{figure}

As we have just seen, the operator $F_A$ creates a domain wall $\partial A$ which separates the two toric code sectors with different values of $s$ when it is applied on the ground state (take the ground state $\vert \Psi^1 \rangle$ for example). The state with a domain wall is no longer a ground state since it has now zero flux and charge excitations. However it is still an eigenstate of the Hamiltonian \eqref{H} with some eigenvalue $E_A$, which is proportional to the length of the path $\partial A$. Now, toric code-like anyons can be created from this state which can be thought of as creating toric code-like excitations in the background of these domain walls. Consider the state $\vert A \rangle=F_A \vert \Psi^1 \rangle$ drawn in figure \ref{domainTC-a}.
A pair of anyons can be created at the end points of a path $\gamma$ by the application of a string operator along this path. Let us consider two distinct cases: the case where $\gamma \cap \partial A=\emptyset$ and $\gamma\cap \partial A \neq \emptyset$. In the first case the path $\gamma$ must be either entirely inside or outside the area $A$, which means it will create a pair of charges either inside or outside the area $A$, but as we already know the quantum number $s$ is different in these two regions. Thus the anyonic excitations created outside will be of the kind $x_1$ while the ones created inside will be of the kind $x_2$ (with $x=e,m, \epsilon$). The figures \ref{domainTC-b} and \ref{domainTC-c} illustrate this fact. However interesting phenomena show up when the path crosses the boundary $\partial A$. We will discuss these next.

\subsection*{Fusing anyons into the domain wall}
 
 If there exist a pair of anyons outside of the area $A$, these two anyons can be freely moved and separated from each other without any energy cost. However if one of the anyons fuses with the zero excitations on the boundary, $\partial A$, it forms a zero excitation on the domain wall. It forms a configuration with one unit of energy less. Note that this is not condensation as in the usual case of physical boundaries where the deconfined excitations of the bulk go to the ground state of the boundary Hamiltonian. In the present case we create an isolated flux along with the domain wall. 
 
\begin{figure}[h!]
\centering
\subfigure[A pair of $m_1$ fluxes localized outside the area $A$.]{
\includegraphics[scale=1]{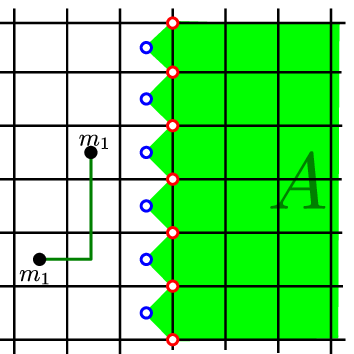}\label{cond-a}
}
\hspace{1cm}
\subfigure[If the flux $m_1$ is moved to a plaquette with a zero flux it fuses into a zero flux excitation.]{
\includegraphics[scale=1]{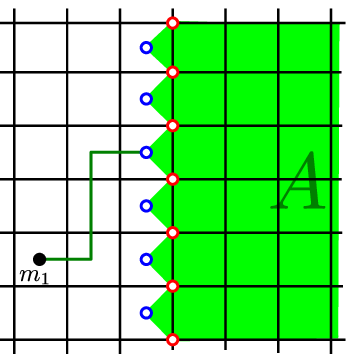}\label{cond-b}
}
\hspace{1cm}
\subfigure[If the flux $m_1$ is moved to a plaquette which is inside the area $A$ it becomes a $m_2$ flux.]{
\includegraphics[scale=1]{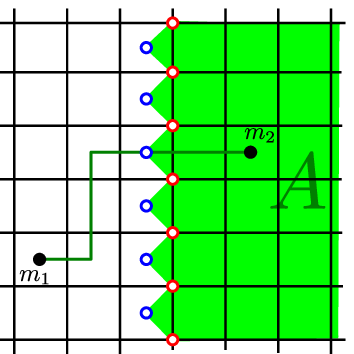}\label{cond-c}
}
\caption{The domain wall can either absorb or scatter the toric code anyonic excitation.}
\label{cond}
\end{figure}

 As an example consider the flux excitations shown in figure \ref{cond-a}. When one of these fluxes is moved to the domain wall and fused with the zero flux excitation as shown in figure \ref{cond-b} it will become a zero flux excitation as well. Thus we get the following fusion rule between the flux $m_s$ ($s\in I$) and a zero flux $m_0$ as 
\beq \label{condflux}m_0\times m_s = m_0 \;, \;\; s\in I\;.\eeq
 A similar thing happens with the deconfined charge excitations, which can be summarized by the following fusion rule 
\beq \label{condcharge}e_0\times e_s = e_0 \;, \;\; s\in I\;.\eeq

In the language of states in the theory, we begin with a broken ground state, $|\psi^1\rangle$ and apply the operator $F_A$ resulting in the state $F_A|\psi^1\rangle$. This state has an energy $E_A$ proportional to the perimeter of the area $A$. The square lattice is now divided into the toric code with $s=1$ outside the region $A$ and the toric code with $s=2$ inside the region $A$. The boundary is the energetic domain wall separating the two sectors. Now apply a string operator creating fluxes denoted by $ X_{\gamma^*} = \bigotimes_{l\in \gamma^*} \Sigma_l^x$ as before. This operator creates a pair of fluxes at the endpoints of $\gamma^*$ in either sector. Let us create it in sector 1 and make the dual string, $\gamma^*$ end at the domain wall. This is the condition for fusing one of the fluxes, $m_1$ into the domain wall. We now have a domain wall excitation $m_0$ as denoted by the fusion rules. 

\subsection*{Synthetic sink}
~

After fusing one of the fluxes from the region with $s=1$ into the domain wall we are left with an isolated flux connected to the domain wall by a string. If we now create a new pair of fluxes in the region with $s=1$, it is easy to see that we can move one of these fluxes and fuse it to the vacuum of $s=1$ to end up in a state similar to the one we started with the difference being in the location of the isolated flux connected to the domain wall by a longer string. This is similar to the presence of a sink for the fluxes. Thus we have synthetically simulated the effect of a sink for the fluxes by creating isolated fluxes connected to the domain walls. It is clear that such effective sinks can be created in any part of the region with $s=1$ by applying the appropriate operators on the domain wall states. 

In a similar fashion we can create sinks for charges using charge excitations instead of flux excitations.
\begin{figure}
	      \begin{center}
		\includegraphics[scale=1.25]{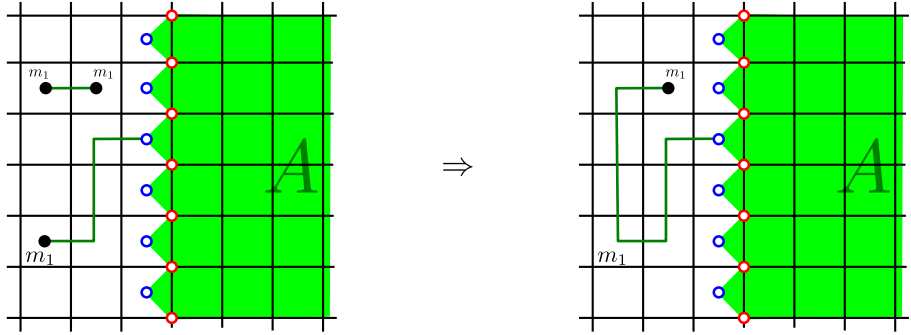}
		\caption{Creation of a sink for the fluxes.}
		\label{condII}
		\end{center}
\end{figure}

\subsection*{The domain wall as an annihilator}
~

We now show how to create a domain wall state whose effect on the deconfined anyons is to annihilate them. We start with a ground state, $|\psi^1\rangle$. We now create a domain wall state by acting with the following operator on a region $A$ :  
$$ D_A = \bigotimes_{j\in A\cup \partial A} G_j = \bigotimes_{j\in A\cup \partial A} \frac{1}{2}\left(\sigma^x\otimes I - \sigma^z\sigma^x\otimes I\right)_j.\footnote{The theory we are working with is a gauge theory and in general the lattice is oriented. Since we are working with the elements of $\mathbb{Z}_2$ whose inverses are itself we need not worry about orientation. In light of this it should be noted that this operator is a valid operator in this gauge theory. Also note that this operator cannot act on the state $|\psi^2\rangle$ as it will kill that state. This is the difference between this operator and the one used previously, namely $F_A$.} $$
Now create a pair of fluxes in the region outside $A$ where $s=1$ using the following string operator : 
$$ M_{\gamma^*} = \bigotimes_{i\in \gamma^*} \frac{1}{2}\left(I\otimes\sigma^x + \sigma^z\otimes\sigma^x\right)_i. $$

The domain wall created by $D_A$ can be seen to act as an annihilator when the string $\gamma^*$ crosses $\partial A$ as can be seen from the identity 
$$ \left(I\otimes\sigma^x + \sigma^z\otimes\sigma^x\right)\times \left(\sigma^x\otimes I - \sigma^z\sigma^x\otimes I\right) = 0.$$
This implies that there is a domain wall that can also annihilate the fluxes in region 1. 

In the language of states this can be written as 
$$ M_{\gamma^*} D_A |\psi^1\rangle = 0, $$
when the dual string crosses the domain wall.

\subsection*{The domain wall as a scatterer}
~

We now show that domain wall created by the operator $D_A$ defined above acts as a scatterer for the fluxes created in the region with $s=2$ states by changing them to the fluxes in the region with $s=1$. For this we create two fluxes with a string operator in the region with $s=2$. To do this we use the following operator defined along a dual string
$$ M'_{\gamma^*} = \bigotimes_{i\in \gamma^*} \frac{1}{2}\left(I\otimes\sigma^x - \sigma^z\otimes\sigma^x\right)_i. $$  

Using the identity 
$$ \left(I\otimes\sigma^x - \sigma^z\otimes\sigma^x\right)\times \left(\sigma^x\otimes I - \sigma^z\sigma^x\otimes I\right) = \left(\sigma^x\otimes I - \sigma^z\sigma^x\otimes I\right)\times \left(I\otimes\sigma^x + \sigma^z\otimes\sigma^x\right)$$
we see that when the string in the region $s=2$ crosses $\partial A$ it gets flipped to a flux in the $s=1$ region. After it crosses the domain wall we make it a string which creates $m_1$, that is $M_{\gamma^*}$.  

The resulting state is the one shown in figure \ref{cond-c}.

\subsection*{A domain wall which permutes anyon types in the two phases}
~

Create a domain region by applying the following operator 
$$ E_A = \bigotimes_{i\in A\cup \partial A}\left[\sigma^x\otimes\sigma^x + \sigma^x\otimes\sigma^z\right]_i.$$
Compared to the earlier cases acting $E_A$ on $|\psi^1\rangle$ results in a different state. In some sense we can think that this dualizes the lattice in region $A$. So the $A_v$ operators in region $A$ now become $B_v$ operators. We show the explicit action on $|\psi^1\rangle$,
$$ E_A|\psi^1\rangle = \prod_{v\notin A} \left(1+A_v\right)\prod_{v\in A}\left(1+B_v\right) \prod_{v\in\partial A} \left(1+ A_v^0\right) |1;~ +1, +1,\cdots, +1\rangle_{l\notin A} |2;~\chi, \cdots, \chi\rangle_{l\in A, \partial A}, $$
with $\chi = |(+1) + (-1)\rangle$.

If we now create a pair of fluxes with a string operator in a region with $s=1$ that also crosses into the region with states in $s=2$, created by the operator $E_A$, we see that this string gets flipped into a string creating charges in the $s=2$ region. That is the operator which excites fluxes in sector 1 will excite charges in sector 2 after crossing the domain wall.

Thus we can create domain walls and domain regions which also switch between anyon types of the two topologically ordered phases.

\section{Generalization to other groups}
~

The local degrees of freedom of the model we have just presented are the elements of $\mathcal{H}_l=\mathbb{C}^2\otimes \mathbb{C}\mathbb{Z}_2$. This space can also be written as a direct sum $\mathcal{H}_l=\mathbb{C}\mathbb{Z}_2\oplus \mathbb{C}\mathbb{Z}_2$, which means this model can be thought of as a direct sum of two toric codes and, in this case, the two phases ($s=1$ and $s=2$) do not represent two different topological phases since they have the same set of anyons satisfying the same fusion rules and braiding statistics. We can generalize the model for the direct sum of two QDMs of two groups $G_1$ and $G_2$, by defining the local Hilbert space as $\mathcal{H}_l=\mathbb{C}G_1\oplus\mathbb{C}G_2$, whose basis is $\{\vert s,g\rangle:s\in I,\; \hbox{and}\;\; g\in G_s\}$. The Hamiltonian is the same as \eqref{H}, with the plaquettes and vertex operator as $B_p=B_p^1+B_p^2$ and $A_v=A_v^1+A_v^2$. But now the operators $B_p^r$ and $A_v^r$ acts as the plaquette and vertex operator of the QDM of the group $G_r$ if all the local state in $\partial p$ and in $\hbox{star}(v)$ has $s=1$, otherwise they project the state to the zero vector. Just as in the $\mathbb{C}\mathbb{Z}_2\oplus \mathbb{C}\mathbb{Z}_2$ case, there are two kinds of ground states which we represented by
$$\vert \Psi^s \rangle=\prod_v \l \mathbb{1}+A_v\r\vert s; e_s,e_s,\cdots,e_s\rangle\;,$$
where $e_s\in G_s$ is the identity of the group $G_s$. Of course there are more states that can be obtained by each of the states above and it is not hard to convince ourselves that this model has one ground state for each ground state of QDM of $G_1$ and another one for each ground state of $G_2$, and therefore its ground state degeneracy is given by $\dim\mathcal{H}^0=\dim\mathcal{H}_1^0+\dim\mathcal{H}_2^0$, where $\dim\mathcal{H}_r^0$ is the ground state degeneracy of the QDM of $G_r$.

From the ground state $\vert \Psi^s \rangle$ one can create all the anyonic excitations of the QDM of $G_s$ by applying the ribbon operators (see \cite{del} for details), and thus all the anyonic excitations of the these two QDMs can be realized in this model. In the $\mathcal{H}_l=\mathbb{C}\mathbb{Z}_2\oplus \mathbb{C}\mathbb{Z}_2$ case we were able to create zero flux and zero charges excitations, corresponding to the states with mismatch of diamond colors in $\partial p$ and/or $\hbox{star}(v)$, here we can create zero flux and charges as well. We just have to apply the operator $F_l$ that flips the diamond color on a link in order to create these kind of excitations, or one could apply this operator in all the links which are inside an area $A$ in order to create a domain wall, as the one shown in figure \ref{domain-c}. It is a known fact \cite{pt} that the anyonic excitations of the QDMs of $G$ (and so its topological phases) are given by irreducible representations of the quantum double algebra $D(G)$. So if one create a domain wall as the shown in figure \ref{domain-c} the excitations which are in the phase $s$ are given by the irreducible representations of $D(G_s)$, but they all can be condensed into the domain wall with the same fusion rules shown in equations \eqref{condflux} and \eqref{condcharge}. Finally, the two topological phases, corresponding to topological phase of QDM($G_1$) and QDM($G_2$) can be simultaneously realized in the same lattice as long as there is a domain wall separating these two phases, and since the groups $G_1$ and $G_2$ do not need to be the same, the two coexisting phases can be different topological phases. As in the earlier case we can annihilate anyons at the domain wall and also switch between anyon types by creating appropriate domain walls.

\section{Remarks}
~

The model we have presented can realize the topological phases of two different QDM of the Kitaev type. Moreover, this model can realize these two topological phases either separately or coexisting as long as there is a domain wall formed of zero fluxes and zero charges separating them. These domain walls are not like the usual ones, as for example the ones where the two dimensional surface where the system lives has a physical boundary \cite{Shor}. They are part of the spectrum of the theory and thus can be created, extended and even destroyed (with an energy cost proportional to the path length), although the excitations living on the domain wall cannot be moved freely and cannot be fused with another excitation living also in the domain wall. Nevertheless they can be fused with the anyons excitations, as shown in equations \eqref{condflux} and \eqref{condcharge}. In other words, the anyon excitations can get condensed when they are moved into the domain wall as show in figure \ref{cond-b}. Therefore, the domain wall is what allows the two different phases to coexist. 

This model can be thought of as a direct sum of two QDMs, it means the degrees of freedom are either elements of the group $G_1$ or $G_2$,  and they are controlled by the addition of an extra quantum number $s\in I$. This construction may seem artificial at first glance, but there is a strong algebraic structure behind it, namely groupoid algebras. Just as the group algebra $\mathbb{C}G$ of a given group $G$ is the main algebraic structure behind the QDM of the Kitaev type, the main structure behind this model is the groupoid algebra $\mathbb{C}\mathcal{G}$, which is very similar to the group algebra apart from the fact that $\mathcal{G}$ is a groupoid instead of a group. Nevertheless the groupoids we are considering are of a special kind, it is formed by two disconnected groups as follows: let $\mathcal{G}= G_1\cup G_2$. The product of two elements $a,b\in \mathcal{G}$ is defined if, and only if, $a,b\in G_s$, and in this case the product is the usual group product $ab\in G_s$. The groupoid algebra $\mathbb{C}\mathcal{G}$, whose basis is $\{ \phi_g: g\in \mathcal{G} \}$ is defined as
$$\phi_a \phi_b = \left\{\begin{array}{ccc} \phi_{ab} &,& \exists ab \in \mathcal{G}\;, \\ 0 &,& \hbox{otherwise.} \end{array}\right.$$
The local Hilbert space $\mathcal{H}_l=\mathbb{C}G_1\oplus\mathbb{C}G_2$ can be defined as $\mathcal{H}_l=\mathbb{C}\mathcal{G}$. Unlike the group algebra, the groupoid algebra is not a Hopf algebra, it is an example of a weak Hopf algebra \cite{dn}. 

Note that the groupoid algebras we have considered are not similar to the direct product of two groups. The zeros in the groupoid algebra is the main difference between the two cases as it does not exist in the definition of the direct product of two groups. Thus if we construct a Kitaev type QDM based on the direct product of two groups we will not see the domain walls that we saw in the QDM based on the groupoid algebra. In \cite{p2} we have showed how to construct the quantum double of a given input using the construction of Kuperberg's invariant \cite{kuper}. We can think of the model in this paper as choosing that input to be that of a groupoid algebra.

A more general construction can be done if one consider $\mathcal{G}$ as being any groupoid. We will report features of these models in an accompanying paper. Some of these features include generating domain walls between a topologically ordered phase containing deconfined excitations and a phase where these excitations are confined with string tension terms. Models with string tension have been considered earlier in~\cite{del, p1}. Quantum double models based on a particular type of quantum groupoids have been considered earlier in~\cite{chang}.

Defects in topologically ordered phases have gained wide attention~\cite{wang, frad}. These papers describe the possible topological order that emerges when one considers defects (or an external symmetry $G$) in the system by studying in detail the theory of $G$ crossed braided extensions of the parent fusion category. However there is still a dearth of exactly solvable models that describe these topologically ordered phases with defects. Some early works include the example provided by Bombin~\cite{Bombin, p2} and simple generalizations of the same~\cite{wen3, wen4}. The theory of defects were studied further in~\cite{maisham}. Tunneling between topologically ordered phases were also studied in~\cite{bais1, bais2}. It will be interesting to see if the models constructed here describe similar phases to the ones studied earlier.

\section{Acknowledgments}
~

The authors would like to thank FAPESP for support during this work. PP thanks Zhenghan Wang for pointing out~\cite{chang}.

\end{document}